\def\aj{\ {AJ}\ }

\def\mnras{\ {MNRAS}\ }

\documentclass[]{elsart}

\newcommand{\be}{\begin{equation}}
\newcommand{\ee}{\end{equation}}
\newcommand{\bea}{\begin{eqnarray}}
\newcommand{\eea}{\end{eqnarray}}
\newcommand{\nobs}{\mbox{$n_{\rm obs}$}}
\newcommand{\nsys}{\mbox{$n_{\rm sys}$}}
\newcommand{\npl}{\mbox{$n_{\rm pl}$}}
\newcommand{\chisqbig}{\mbox{$\chi^2_{\rm big}$}}
\newcommand{\chisqsmall}{\mbox{$\chi^2_{\rm small}$}}
\newcommand{\go}{\ {\raise-.5ex\hbox{$\buildrel>\over\sim$}}\ }
\newcommand{\lo}{\ {\raise-.5ex\hbox{$\buildrel<\over\sim$}}\ }
\newcommand{\mps}{\ensuremath{{\rm m}\,{\rm s}^{-1}}}

\begin{document}

\begin{frontmatter}

\title{Parallel Algorithm for Solving Kepler's Equation on Graphics Processing Units:\\
\vspace{15pt}
Application to Analysis of Doppler Exoplanet Searches}

\author[UF]{Eric B. Ford}
\address[UF]{Department of Astronomy, University of Florida, 211 Bryant Space Science Center, PO Box 112055, Gainesville, FL, 32611-2055, USA}

\begin{abstract}
We present the results of a highly parallel Kepler equation solver
using the Graphics Processing Unit (GPU) on a commercial nVidia
GeForce 280GTX and the ``Compute Unified Device Architecture'' (CUDA)
programming environment.  We apply this to evaluate a goodness-of-fit
statistic (e.g., $\chi^2$) for Doppler observations of stars
potentially harboring multiple planetary companions (assuming
negligible planet-planet interactions).  Given the high-dimensionality
of the model parameter space (at least five dimensions per planet), a
global search is extremely computationally demanding.  We expect that
the underlying Kepler solver and model evaluator will be combined
with a wide variety of more sophisticated algorithms to provide
efficient global search, parameter estimation, model comparison, and
adaptive experimental design for radial velocity and/or astrometric
planet searches.

We tested multiple implementations using single precision, double
precision, pairs of single precision, and mixed precision arithmetic.
We find that the vast majority of computations can be performed using
single precision arithmetic, with selective use of compensated
summation for increased precision.  However, standard single precision
is not adequate for calculating the mean anomaly from the time of
observation and orbital period when evaluating the goodness-of-fit for
real planetary systems and observational data sets.  Using all double
precision, our GPU code outperforms a similar code using a modern CPU
by a factor of over 60.  Using mixed-precision, our GPU code provides
a speed-up factor of over 600, when evaluating $\nsys > 1024$ models
planetary systems each containing $\npl = 4$ planets and assuming
\nobs=256 observations of each system.  We conclude that modern GPUs
also offer a powerful tool for repeatedly evaluating Kepler's equation
and a goodness-of-fit statistic for orbital models when presented with a large
parameter space.
\end{abstract}

\begin{keyword}
  gravitation --
  planetary systems --
  methods: numerical -- 
  techniques: radial velocities
\end{keyword}
\end{frontmatter}

\section{Introduction}
\label{Sec:Intro}
In the classical bound two-body problem, a body follows an elliptical
orbit.  While there are several possible parameterizations, one
particularly useful parameterization identifies the instantaneous
location of a body along its orbit by the {\em eccentric anomaly}, the
angle between the pericenter direction and the current location,
measured from the center of the ellipse.  According to Kepler's second
law, the body travels more quickly near the pericenter, so the
eccentric anomaly does not increase uniformly.  Instead, the {\em mean
anomaly} is defined to increase uniformly in time, and {\em Kepler's
equation} relates the eccentric anomaly, $E$, to the mean anomaly,
$M$, by
\be
M = E - e \sin E,
\ee
where $e$ is the orbital eccentricity.  As Kepler's equation is
transcendental, there is no explicit solution in terms of elementary
functions.  Given the fundamental importance and extremely frequent
application of Kepler's equation, numerous
mathematicians and scientists have developed many methods for solving
it numerically,
typically involving infinite series or iterative solutions.  
In this paper we combine an existing iterative algorithm with the
computing capabilities of a modern {\em graphics processing unit}
(GPU) to develop an extremely efficient method for repeatedly solving
Kepler's equation.  While this method has many potential applications,
we chose to focus on an application to modern searches for extrasolar
planetary systems.

\subsection{Motivation for Research in GPU Computing}

Modern GPUs are extremely powerful and capable processors.  The peak
computational capabilities of currently available graphics cards
significantly exceeds that of top-of-the-line central processing units
(CPUs).  For example, a modern, high-end, quad-core CPU achieves
roughly 81 billion floating point operations per second (FLOPS) on
``linpack'', a relatively easy benchmark for CPUs.  In contrast, a
single GT200 chip offers a theoretical peak performance of over 1
trillion FLOPS.  However, harnessing this power poses a significant
challenge, as numerical algorithms and memory access patterns that are
well-suited for CPUs are typically highly inefficient on modern
highly-parallel GPUs.  The performance of GPUs already exceeds that of
modern  CPUs by one or two orders of magnitude for selected
applications which are well suited to the
architecture of GPUs.  In addition to raw computing performance, GPUs
offer similarly impressive metrics for performance per dollar and
performance per watt.

``Moore's law'' describes the historical exponential increase in the
number of transistors within an integrated circuit and is often
used to extrapolate the performance of computers in coming
years. During the past several decades, CPUs have made dramatic
improvements in computational performance, both by reducing the
size of etchings and increasing the clock speed.  In recent years, CPU
designers have shifted their attention from improving the speed of
serial operations to making processors capable of performing more
computations in parallel.  High-end CPUs are now avaliable with four
or eight processor ``cores'' and the trend towards multi-core
computing is expected to accelerate in years to come.  GPUs achieve
their high performance by devoting a larger fraction of the microchip
to processors and a smaller fraction of the chip to logic and caching.
Current high-end GPUs include over a hundred floating point units that
operate using the shared instruction multiple data (SIMD) paradigm.
Clearly, this raises new issues for programming and not all algorithms
can be implemented efficiently on such a highly parallel architecture.
Yet, given the direction of computing hardware, we expect there will
be a renewed focus on the development of algorithms for
highly-parallel multi-core hardware.


With traditional CPUs, the speed of an numerical algorithm is largely
determined by the number of FLOPS required.
However, even the current generation of GPUs have such high compute
throughput that they are often limited by memory access.  Thus,
traditional algorithms chosen to minimize floating point operations
may be less efficient than alternative algorithms that are optimized
to minimize memory requirements.  For example, an iterative algorithm
may require more floating point operations to converge to a desired
precision than an explicit algorithm, causing it to run more slowly
on a traditional CPU.  Given the memory access patterns of GPUs, the
iterative algorithm may be able to perform the extra computations at
very little marginal cost, thereby providing higher performance.
Recent developments among semi-conductor manufacturing companies
suggest that highly-parallel multi-core processors will soon become
the norm for scientific computation.  Thus, we anticipate that our
algorithms for highly-parallel solving of Kepler's equation and evaluating
goodness-of-fit statistics will have enduring
value for future generations of GPUs, CPUs, and other flavors of
many-core processors (e.g., FPGAs).  

\subsection{Recent Development in GPU Computing for Scientific Applications}

Several developments have enabled GPUs to become a powerful tool for
scientific applications.
First, the performance of GPUs is increasing much more rapidly than
that of CPUs, thanks to their highly parallel design of GPU.  The mass
production of GPUs for gamers results in very attractive prices,
similar to the way that mass market PCs have led to the development of
affordable high performance CPUs.
Second, early generations of graphics cards were very specialized and
not well suited for general purpose scientific computing.  The two most
recent generations of graphics use a unified architecture
that is much more amenable to scientific computing.  Our research is
greatly aided by GPU manufactures developing programming languages that make it much easier to perform
general-purpose floating point calculations with high-end GPUs.  In
particular, nVidia has recently released a Compute Unified Device
Architecture (CUDA) compiler and software development kit that makes
it much easier to program GPUs using a programming language very much
like C, but with some extensions and an additional library of
functions to access the GPU.  Importantly, nVidia promises CUDA code will
be able to execute on future generations of GPUs with a simple
recompiled (though some parameter optimization will be needed to
achieve maximum performance; nVidia 2008).  Indeed, the code developed for this
project was originally written for a G80-based GPU.  With only a
trivial recompile, the code worked on the recently released
GT200 GPU and provided an even more substantial speed-up.
Finally, the most recent generation of GPUs contains hardware for
performing double precision arithmetic.  This facilitates the
application to a much wider range of scientific problems.
Unfortunately, the GT200 is only capable of performing an eighth as
many operations at once when performing double precision calculations
as compared to performing single precision arithmetic.  Thus, the
higher precision comes at a considerable cost.  Therefore, we
investigate the use of mixed precision arithmetic, as well as
techniques such as compensated summation that use multiple single
precision values to represent a number with increased precision.

\subsection{Relation to Previous Research}

Recently, computer scientists have implemented efficient algorithms
for a few common numerical methods (e.g., Fast Fourier Transform
(FFT), Basic Linear Algebra Subprograms (BLAS)) and provided libraries
of functions (or application programming interface; API) to facilitate
scientists harnessing the power of GPUs to perform such computations
much more rapidly by calling these routines from traditional
high-level programming languages such as C/C++, Fortran, Matlab,
python, and IDL.

Astrophysicists have also begun to harness the power of GPUs.  For
example, Portegies Zwart et al.\ (2007) developed a GPU kernel to
perform the force calculation step for an n-body system containing of
$N\sim~10^4$---$10^6$ bodies and achieved speed-up factors of upto
$\sim~20$.  Hamada \& Iitaka (2007) and Belleman et al.\ (2008)
developed similar routines that achieved a speed-up factor of
$\sim~100$ for $N\sim10^5$. Nyland et al.\ (2007) have released an
even faster demonstration code that performs both the force
calculation and a direct n-body integration code as part of nVidia's
CUDA SDK.  While this results in impressive computer graphics, it not
sufficient accuracy for scientific applications, since it uses an
Euler integrator.  Moore \& Quillen (2008) have developed a
science-grade n-body code optimized for planetary dynamics.  Other
astrophysics-related applications include FFTs and data processing for
radio astronomy (e.g., Harris et al. 2008).

Here we describe a simple code for solving Kepler's equation on a GPU
using CUDA.  The GPU kernel code can be applied to a variety of
problems, and we choose to demonstrate it's application to evaluating
orbital models for the radial velocity variations of a star perturbed
by one or more planets.  In this case, we implement the entire
calculation on the GPU, providing for parallel data reduction and
reducing the amount of data that needs to be transferred between the
CPU and GPU.  

We find that for our application, single precision arithmetic is
sufficient when solving Kepler's equation and for most other
calculations.  However, higher precision is needed for at least one
step to achieve science-grade accuracy.  Thus, we implemented the
model evaluation using a combination of single precision, double
precision, and pair of single precision arithmetic.  Our algorithm for
evaluating $\chi^2$ provides an impressive speed-up factor of over 600
relative to performing the same calculation on a top-of-the-line CPU
(2.6GHz AMD Opteron 2218).  

In \S\ref{Sec:Eqns} we describe the physical model and the
equations to be solved.  We provide a brief overview of the
architecture and programming environment of a modern GPU in
\S\ref{Sec:Cuda}.  We describe our basic implementation in
\S\ref{Sec:Implementation} and discuss the accuracy and performance
of our code with various optimizations in \S\ref{Sec:Optimizations}.
Finally, we discuss the implications and future prospects in
\S\ref{Sec:Discussion}.

\section{Basic Equations}\label{Sec:Eqns}

We consider the problem of global search to identify plausible orbital
solutions for an extrasolar planetary system based on radial velocity
observations of the host star.  We desire to evaluate a
goodness-of-fit statistic (e.g., $\chi^2$) for each of many orbital
models.

\subsection{Physics}\label{Sec:Physics}

We focus on planetary systems for which any planet-planet
interactions are either small or occur on sufficiently long
timescales that they can be ignored.  In this case, the radial
velocity perturbation on the host star ($\Delta v(t_k)$) 
can be written as the linear
sum of the perturbations due to each planet ($\Delta v_{j}(t_k)$) at time $t_k$,
\be
\Delta v(t_k) = \sum_j^{\npl} \Delta v_j(t_k).
\ee
Since each planet is assumed to travel on a Keplerian orbit,
\be \label{Eqn:Rv}
\Delta v_j(t_k) = \frac{K'_j}{1-e_j\cos E_{jk}}\left[ -\cos \omega_j \sin E_{jk} + \sqrt{(1-e_j)(1+e_j)} \sin \omega_j \cos E_{jk}  \right]
\ee
where $e_j$ is the planet's eccentricity, $\omega_j$ is the planet's
eccentric anomaly, $E_{jk}$ is the planet's eccentric anomaly, and
$K'_j$ is closely related to the radial velocity amplitude of the
planet's perturbation (Danby 1988).  Note that for computationally
purposes we use $K' \equiv K/\sqrt{1-e^2}$, where $K$ is the standard
radial velocity amplitude and both are related to the planet's mass,
orbital period, eccentricity, and inclination.  The eccentric anomaly
is related to the time of the observation via the mean anomaly
($M_{jk}$) and Kepler's equation (Eqn.\ 1).  The mean anomaly
($M_{j}$) is given by
\be
M_j(t) = 2\pi t/P_j + M_{0,j},
\ee
where $t$ is the time, $P_j$ is the orbital period, and $M_{0,j}$ is the
mean anomaly at the chosen epoch ($t=0$).

\subsection{Model Evaluation}\label{Sec:ChiSq}

We compute a goodness-of-fit statistic for each model and demonstrate
our method using the common $\chi^2$ statistic, defined by
\be \label{Eqn:ChiSq}
\chi^2 = \sum_k^{\nobs} \frac{\left(\Delta v(t_k) - v_k\right)^2}{\sigma_j^2+\sigma_k^2},
\ee
where $v_k$ is the observed radial velocity perturbation at time
$t_k$, $\sigma_k$ is the associated measurement uncertainty, and $\sigma_j$ is the {\em jitter} parameter that can
account for additional noise due to either astrophysical or
observational effects.  For simplicity, we set $\sigma_j=0$\mps, but
we include it as a variable in our code to provide realistic
benchmarks.

\section{nVidia Architecture \& CUDA}\label{Sec:Cuda}

While GPU hardware is very powerful, they have more complicated memory
hierarchies that make them more difficult to program and create a need
for algorithms that are optimized for parallel execution. For example,
on the nVidia GT200 GPU, there are 30 {\em multi-processor units}
(MPUs), each containing 8 standard {\em execution pipes} (EPs) that
perform simple numerical calculations (integer or single precision
addition, multiplication, or multiple-and-add in one clock cycle).
Each MPU contains two {\em super function unit} (SPU) pipes capable of
performing single precision operations such as reciprocal, reciprocal
square root, and trig functions, and one double precision EP.  Each
MPU can apply these three pipes by performing SIMD (shared instruction
multiple data) operations, meaning that each EP must perform the same
mathematical operations, but can perform them on many different sets
of different numerical values.  Each data set being operated is
represented by a {\em thread} and each MPU can interleave the
execution of upto 768 active threads to maximize performance.
Fortunately, this design is well suited to solving Kepler's equation
and/or evaluating a model goodness-of-fit statistics many times.

Due to the impressive computational capabilities of GPUs, real-world
performance can be limited by the need to access memory.  In
particular, traditional algorithms that minimize computation at the
expense of storing many values in memory may result in GPUs sitting
idle while waiting for data to be read from memory.  To ensure that
the EPs are utilized efficiently, the threads are grouped in {\em
blocks} ($\ge$64 threads/block), each executed on a single MPU.  The
MPU can interleave operations for different threads in one or more
execution blocks to keep each EP occupied with calculations.  All
threads can access a large amount (1GB) of {\em device memory} that is
shared between all MPUs and threads.  While this memory
has a high latency (hundreds of clock cycles), it is still
$\sim$5-10$\times$ faster than a CPU accessing non-cached memory using
state-of-the-art DDR3 RAM.  To reduce memory access latency,
programmers can place some data in either a {\em constant memory}
cache or a {\em texture memory} cache, each of which essentially
provides a low latency read-only cache (optimized for different access
patterns) that is shared between all the MPUs.  Each MPU
also has it's own low-latency read-write {\em shared-memory} that can
be accessed by all threads within a thread block.  In a simplistic
parallelization scheme of one Kepler's equation per thread, there is
no need for communication between threads, so the shared-memory can be
used as additional low-latency memory cache that is divided evenly
between all threads in a block (that are being executed on a single
MPU).  Finally, each EP has a set of low-latency
read-write {\em registers} that can not be used to communicate between
threads.  We perform our calculations using an nVidia GT200 GPU that
contains 30 MPUs, each with 16 kB of shared memory, 8kB of constant
cache, and 8kb of texture cache.  Each EP has 8kB of registers.

At the beginning and end of a scientific calculation, data must be
transferred between the device {\em global memory} and standard memory
connected to the host CPU.  For some algorithms that require large
amounts of input/output and a modest amount of computation, this
transfer can become significant.  Therefore, it is often advantageous
to perform basic data reduction on the GPU.  For example, if we were
to transfer the every solution to Kepler's equation, then nearly two
thirds of the total time would be devoted to the GPU to host transfer of
results (see \S\ref{Sec:Communications}).  In our case, we use the
solutions to Kepler's equation to evaluate the predicted velocities of
the orbital period, compare them to a set of observations, and
evaluate $\chi^2$.  Since we only need to return one value of $\chi^2$
for each system, over 92\% of the wall clock time is devoted to
computation.

\section{Implementation}\label{Sec:Implementation}

We describe a code that evaluates $\chi^2$ for a set of radial
velocity observations and a large number of orbital models.  While the
mathematical operations are straightforward, there are multiple
possible mappings of the mathematical operations to the hardware and
memory architecture of a GPU.  We describe two basic implementations
below. 

\subsection{CPU-GPU Communications}\label{Sec:Communications}

In each implementation, the first step is to transfer the
observational data and model parameters from the host to the GPU.
This includes a list of \nobs observation times ($t_k$), observed
velocities ($v_k$), and measurement uncertainties ($\sigma_k$).  We
consider these values to be constants to be used for evaluating all
models.  In \S\ref{Sec:OptCache}, we evaluate the potential
benefits of caching this data.  Second, we
transfer the list of model parameters.  The radial velocity
perturbation of each planet is described by five parameters
$\left\{P,K,e,\omega,M_0\right\}$, so we must transfer a list of \nsys
$\times$ \npl $\times$ 5 floating-point values from the host to the GPU,
where \nsys\ is the number of planetary system models to be evaluated
and \npl\ is the number of planets per system.  At the end of each
calculation, we transfer the value of $\chi^2$ for each of the \nsys\ 
orbital models from the GPU to the host.  

\subsection{Mean Anomalies}\label{Sec:MeanAnomaly}

Once the observational data and model parameters have been transferred
to the GPU, the first step is to generate a list of $\nsys \times
\npl \times \nobs$ mean anomalies, one for each planet of each model
at each observation time.  In order to reduce the number of times the
orbital periods are read from global memory, we use only \nsys $\times$
\npl threads, where each thread solves for the mean anomaly of one planet at
each of the \nobs observations.  If $i$ indexes the systems, then the
mean anomalies are given by
\be \label{Eqn:MeanAnom}
M_{ijk} = 2\pi \left[ t_k/P_{ij} - \mathrm{floor}(t_k/P_{ij}) \right] - M_{0,ij},
\ee
where $\mathrm{floor}(x)$ is the near integer less than or equal to
$x$.  We will show in \S\ref{Sec:OptPrecission} that single precision
provides sufficient precision for most operations, with the notable
exception of the calculation inside the square brackets.  Therefore,
we assign the computation in square brackets to a separate function to
be executed in double precision.  The multiplication by $2\pi$ and
subtraction of the mean anomaly at epoch (as well as an if-add
statement to ensure that the mean anomaly lies in [0,$2\pi$) ) can be
performed in single precision just before solving Kepler's equation.

\subsection{Kepler's Equation}\label{Sec:Kepler}

We solve Kepler's equation iteratively using the Newton-Raphson method generalized
to provide quartic convergence. Following Danby (1988), we define
\be
f(E) = E - e \sin E - M
\ee
and let $E_{l+1} = E_j+\delta_{l3}$, where
\bea
\delta_{l1} & = & -f_l/f'_l \\
\delta_{l2} & = & -\frac{f_l}{f'_l+\delta_{l1}f''_l/2} \\
\delta_{l3} & = & -\frac{f_l}{f'_l+\delta_{l1}(f''_l+\delta_{l2}^2f'''_l/3)/2}.
\eea
We evaluate $\sin(E)$ and $\cos(E)$ simultaneously via the $\mathrm{sincos}$ function,
 so the derivatives
of $f(E)$ come at virtually no extra computational expense.  An
initial guess of $E_0 = M + 0.85 e
\times\mathrm{sign}(\sin M)$ results in excellent convergence within a
few iterations for nearly all $e$ and $M$.  We assume that $M\in
[0,2\pi)$, so we use a single {\tt if} statement to decide the sign.
We stop iterating once $\left|f(E)\right|<10^{-6}$.  We observe no
significant effects of stopping the Kepler solver after five
iterations, but also no significant performance penalty for allowing
arbitrarily greater number of iterations.  While removing the test for
maximum number of iterations provides a slight performance boost
($\simeq~2.3\%$), we opt to maintain a guaranteed maximum number of
iterations to ensure our code does not trap the GPU in an infinite
loop.  Five iterations provided acceptable accuracy for eccentricities
$e<0.99$, but we found allowing for a maximum of ten iterations
resulted in no noticeable performance penalty.  When calculating
$E_{ijk}$, we use one thread per planet, where each thread loops over
$\nobs$\ observations, so as to avoid repeatedly reading the
eccentricities from global memory.

\subsection{Radial Velocity \& Model Evaluation}\label{Sec:RvEval}

We evaluate the radial velocity perturbation of each planet at
each time ($\Delta
v_{ijk}$) according to Eqn.\ \ref{Eqn:Rv}.  We use the
fast version of the reciprocal square root and calculate the argument
as $1-e^2=(1+e)(1-e)$ to limit round-off error for low $e$.  

\label{Sec:ChiSqEval}

We evaluate $\chi^2_i$ for each of \nsys\ sets of model parameters
using Eqn.\ \ref{Eqn:ChiSq}.  We find that for the vast majority of
data sets and model parameters, standard single precision arithmetic
is sufficient.  However, to provide more generality, we use a pair of
single precision floating point values and compensated summary in
order to improve the precision of $\Delta v_{ik}$ and $\chi^2$.  That is the value of
$\chi^2$ is represented by the sum of \chisqbig and \chisqsmall.  We
calculate $\chi^2=\sum_k \chi^2_k$ by accumulating the single
precision values $\chi^2_k$.  For each $k$, we first calculate
temporary variables $u =
\chi^2_k-\chisqsmall$ and $w = \chisqbig+u$.  Then, we update
$\chisqsmall = (w-\chisqbig)-\chi^2_k$ and $\chisqbig = u$ (Kahan
1965).

We consider two implementations of the radial velocity and model
evaluation: 1) the radial velocity perturbation is evaluated by one
function (one thread per planet) that writes each $\Delta v_{ijk}$ to global device memory
while the model evaluation is calculated by a separate function (one thread per system), and
2) both the radial velocity perturbations and the model evaluation are
performed by a single function (one thread per system).  In the first implementation,
evaluating the radial velocity perturbations requires $\nsys \times
\npl \times (4+ \nobs)$ values to be read from device memory and
$\nsys \times \nobs \times \npl$ values of the radial velocity
perturbation from each planet to be written to device memory and then
read back from device memory by the model evaluation function.  The
second implementation requires $5 \times
\nsys \times \nobs \times \npl$ values to be read from device
memory to calculate the radial velocity perturbations.  The above
numbers do not include the observation times, observed radial
velocities or measurement errors, which are the each implementation.
These memory accesses are negligible since they can be amortized over
all threads within a block via cooperative loading and cached in
either constant cache or shared memory (see \S\ref{Sec:OptCache}).
The memory requirements of each function are summarized in Table 1.


\section{Optimizations}\label{Sec:Optimizations}

We have attempted a number of optimizations in order to
improve performance.  As a benchmark, we evaluate 122,800 models, each
containing 4-planets and using 256 observations.  The values of the
eccentricity and mean anomaly at epoch are drawn from uniform distributions
over the intervals [0,0.99) and [0,2$\pi$).  The accuracy and performance
benchmarks are summarized in Table \ref{Tab:Performance}.  We discuss
each of several optimizations below.

\subsubsection{Mixed Precision Arithmetic}\label{Sec:OptPrecission}
We found the biggest performance increase to come from replacing
double precision arithmetic with single or mixed precision arithmetic.
If all calculations are performed in double precision, then the GPU
can only use one eighth as many EPs and most arithmetic operations
take multiple clock cycles.  We found that a purely double precision
GPU outperformed a single-threaded CPU implementation by a factor of
roughly 20.  In the double precision implementation,the GPU spent 90\%
of the time solving Kepler's equation.  However, many applications do
not need to solve Kepler's equation to double precision.  Indeed, for
the purposes of comparing orbital models to radial velocity
observations, we can use single precision for solving Kepler's
equation, calculating radial velocity perturbations, and evaluating
$\chi^2$.  The operation most sensitive to round-off error is
determining the orbital phase.  Therefore, we store the observation
times and orbital periods as double precision values and calculate the
term in the square brackets of Eqn. \ref{Eqn:MeanAnom} using full
double precision.  However, we store the resulting mean anomaly in
single precision and perform the remainder of the calculations using
single precision arithmetic.  This reduced the worst case fractional
error in $\chi^2$ from one part in ten (using pure single precision)
to one part in $10^4$ (using mixed precision).  We found that using
mixed precision accelerated the solving of Kepler's equation by a
factor of $\simeq~23$ and provided an overall performance speed-up of
$\simeq~11$ when calculating $\chi^2$ (relative to the same
calculation using purely double precision arithmetic on the GPU; see
\S\ref{Sec:OptTransfer}).

The other arithmetic operations where roundoff error could be
significant are the summation of the radial velocity perturbations
from multiple bodies and the summation of $\chi^2$.  The first is not
an issue for the planetary case, since even very-short-period giant
planets result in a radial velocity perturbation of only hundreds of
meters per second and the Doppler measurement precision is of order a
meter per second or more.  However, single precision ($\simeq~2\times
10^{-7}$) might be limiting for planets in binary star system and/or
other planet detection techniques with larger dynamic range (e.g.,
astrometry, pulsar timing, transit timing variations).  Thus, we
accumulate the radial velocity perturbations using a pair of single
precision numbers and the Kahan algorithm for compensated summation (see \S\ref{Sec:ChiSqEval}).
Similarly, we found that the roundoff error in $\chi^2$ can become
significant (of order unity) for extremely poor orbital models.  Since
such models are typically discarded or given an extremely small
weight, this is not a serious issue for our applications.  However, we
recognize that this could become significant for future very large
data sets.  Applying compensated summation for the radial
velocities and $\chi^2$ values results in less than a 0.1\%
performance penalty.  Therefore, we use a pair of single precision values
and compensated summation to calculate $\Delta v_{ik}$ and $\chi_i^2$ (see
\S\ref{Sec:RvEval}).  


\subsubsection{CPU-GPU Memory Transfer}\label{Sec:OptTransfer}
The second very significant performance enhancement came from
increasing the amount of work assigned to the GPU (i.e., evaluating
predicted radial velocities at each time and $\chi^2$ for each model),
so that only one value ($\chi^2$) per system needed to be transferred
back to the CPU's host memory.  In our initial mixed-precision
implementation, we applied the GPU to solving Kepler's equation only,
in which case over two thirds of the total wall clock time was
dominated by the transfer of $\nsys \times \npl \times \nobs$
solutions of Kepler's equation from the GPU to the host memory.  By
implementing the radial velocity evaluation and $\chi^2$ calculation
on the GPU, we reduced the transfer time to $\simeq~9\%$ of the wall
clock time.  We further reduced transfer time using {\em pinned
memory} buffers on the host, so that the total transfer time was less
than $\simeq~7\%$ of the wall clock time.  Finally, we note that our
GPU is connected to the motherboard by a PCI Express 4x slot (maximum
one-way bandwidth of 1000 MB/s), while the graphics card contains a
PCI Express 16x interface.  Thus, we expect that the transfer time
could be significantly reduced if the graphics card were connected to
the motherboard using PCI Express 16x.

\subsubsection{Device Memory Caching}\label{Sec:OptCache}
As a third optimization, we placing the observation times, observed
radial velocities and uncertainties in the constant memory cache
and/or shared memory.  When using the shared memory, we make use of
the {\em cooperative load} strategy described in the CUDA SDK (nVidia 2008).  
For our first implementation (separate functions for radial velocities and
$\chi^2$), we find modest performance gains of 1.5\% by placing the
observed velocities and uncertainties in either the constant chance or
the shared memory.  Using the constant cache for the observation times
was not advantagous, but placing them in shared memory provided a
0.8\% speed up.  Finally, we note that it is important for the array
of mean anomalies at each time to be stored so that different planets
at the same time are adjacent in memory, as this allows reads from
device memory to be coalesced.  Using the reverse ordering resulted
in over four times worse performance.

For our second implementation (single function for evaluating radial
velocities and $\chi^2$), we again find a 0.8\% speed up by placing the
observed times in shared memory, and a slightly greater benefit
(2.5\% speedup) by placing observed velocities and uncertainties in
the constant cache or shared memory.  The first implementation is
$\simeq 20\%$ faster, but requires using a significant amount of
global memory ($16\times\nsys\times\npl\times\nobs$ bytes).

For the number of observations that we considered (256), the number of
active threads per MPU is limited by the number of registers per MPU,
and not by shared memory, even when all observations are cached into
the shared memory.  As long as this is the case, it is
slightly advantageous to use the shared memory as a cache for the
observational data rather than the constant cache.  If the number of
observations were to increase significantly, then it would likely
become preferable to use the constant cache, so as not to limit the
number of blocks and hence threads per MPU.  Similarly, if one were to
implement a more complex algorithm on the GPU (e.g., genetic
algorithm, Markov chain Monte Carlo), then the other code might easily
score a greater benefit by utilizing the shared memory, so that the
combined code would be more efficient if the observational data were
placed in the constant cache.

\subsection{Texture Memory for Initial Eccentricity Guess}\label{Sec:OptTexture}

Next, we attempted to improve performance by using the texture memory
cache to provide a superior initial guess for the Kepler solver.  We
solved Kepler's equation using the CPU at each point on a 2-d grid (16
eccentricities and 96 mean anomalies requiring 6kB of texture cache)
and the results were written to the texture cache.  The initial guess
for the GPU's Kepler solver was based on bilinear interpolation.  We
found a slight decrease in performance for our benchmark case (random
distribution of eccentricities and mean anomalies) and a slight
increase in performance if we used large eccentricities and mean anomalies
that were tightly clustered.  Given the differences were only
$\simeq~0.2\%$, we do not consider this to have been a worthwhile
optimization.

\subsubsection{GPU Occupancy \& Hiding Memory Latency} \label{Sec:OptTuning}
The GT200 GPU contains 240 single precision EPs.  In light of the
latency incurred upon requests to access to the GPU's global memory,
it is important to run many more threads in parallel in order to
maintain a high occupancy of the GPU's EPs.  While CPUs generally hide
memory latency be clever caching, GPUs hide memory latency by having a
large number of active threads that can execute while waiting for
memory reads.  We found that performance generally maximized for the
256 threads per block and 120 or 60 blocks for GPU functions that
calculate the mean anomaly and the radial velocity perturbation by
each planet.  However, performance nearly plateaus once we reach a
size of 128 threads and 60 blocks.  For the function that calculates
both the radial velocity perturbations and the model $\chi^2$ or the
function calculates $\chi^2$ using the previously calculated list of
radial velocity perturbations from each planet, it was advantageous to
use only 15 blocks (one per MPU) and 512 threads per block (the
maximum number of threads currently allowed).  Since this function
performs relatively little arithmetic, the large number of active
threads helps minimize the effects of memory latency and only using a
single block avoids repeatedly caching the observational data.

\section{Discussion}\label{Sec:Discussion}

%
We have implemented a highly parallel Kepler equation solver for GPUs
using the CUDA programming environment.  The nVidia GT200 GPU offers
an impressive speed-up factor of $\simeq 1200$ (55) for solving
Kepler's equation in single (double) precision, relative to a similar
single-threaded algorithm running on a 2.6 GHz AMD Opteron CPU.
However, the time required to transfer data between the CPU and GPU is
quite significant.  Therefore, we recommend implementing further data
reduction on the GPU when practical.  To demonstrate this technique,
we considered the case of evaluating the $\chi^2$ goodness-of-fit
statistic for four planet orbital models using an actual set of radial
velocity observations.  By implementing the Kepler equation solver,
radial velocity evaluation, and model evaluation steps on the GPU, we
significantly reduce the communications overhead to less than 8\% of
the wall clock time.  We find that the vast majority of computations
can be performed in single precision while maintaining an accuracy of
one part in $\simeq~10^{4}$ in $\chi^2$.  However, the calculation of the
mean anomaly of each planet at each observation time must be performed
in double precision to maintain accuracy.  Therefore, we implement
our GPU-based model evaluation using mixed-precision.

We find that using the GT200 the full model evaluation can be
accelerated by a factor of $\simeq 625$ (68) relative to a CPU
implementation using mixed (all double) precision, for $\npl=4$,
$\nobs=256$, and large $\nsys$.  We also provide a second
implementation that requires much less memory and still achieves 86\%
of the performance of our leading implementation.  This low-memory
version may prove useful when combined with more sophisticated
algorithms that may need significant device memory themselves.

%
The use of GPUs for comparing orbital models to radial velocity
observations has the potential to significantly accelerate global
searches for multiple planet solutions.  
A demonstration version
of our code is avaliable online at
\verb+http://www.astro.ufl.edu/~eford/code/cuda_kepler/+.
For the sake of clarity, we present only a simple Monte Carlo code for
$\chi^2$ evaluation.  Clearly, we intended for this GPU kernel to be
combined with more sophisticated algorithms, most likely implemented
in traditional programming languages on the CPU.
Obviously, brute force search and genetic algorithms are extremely
well suited for highly parallel model evaluation.  Iterative
algorithms that require values of the first and/or second derivatives
of the goodness-of-fit statistics might benefit from a similar
algorithm to evaluate partial derivatives in parallel.

We expect that our GPU-based algorithms will be particularly
advantageous for Bayesian algorithms that are increasingly being used
to analyze exoplanet observations.  With standard Markov chain Monte
Carlo (MCMC), one generates several long chains (Ford 2005), making
efficiently parallelization on many-core architectures challenging.
The situation improves somewhat, when one considers parallel tempering
MCMC, as one typically uses several dozen Markov chains for a single
system (Gregory 2007).  We expect that the availability of GPUs for
rapidly evaluating tens of thousands of models will shift attention to
population MCMC algorithms that are naturally suited for highly
parallel architectures.  For example, differential evolution MCMC
(DEMCMC) can apply hundreds or thousands of chains to efficiently
sample from complex and high-dimensional posterior densities (ter
Braak 2006).  Indeed, we have recently implemented a simple DEMCMC
code and found that this can significantly accelerate converge
relative to standard MCMC algorithms, even on CPUs.  We expect that
the combination of DEMCMC and our GPU kernels will be particularly
powerful for efficiently performing Bayesian parameter estimation.  In
principle, this could also be combined with parallel tempering to
empower Bayesian model comparison.  GPU-based model evaluation could
also significantly accelerate other algorithms for Bayesian model
comparison, such as importance sampling or the ratio estimator (Ford
\& Gregory 2006).  Finally, our algorithms could be easily extended to
calculate predictive distributions to be enable practical Bayesian adaptive
experimental design (Loredo 2004; Ford 2008).  Such algorithms may be
particularly relevant for the efficient scheduling of observations by
large space missions, such as the Space Interferometry Mission and/or
Terrestrial Planet Finder, as they search for terrestrial-mass planets
in the habitable zone of nearby stars.

%

\section*{Acknowledgments}
This research was supported by UF, the Research Opportunity Incentive
Seed Fund, and NASA JPL subcontract \#1349281.  
We thank Jainwei Gao, Mario Juric, Jorg Peters, Alice Quillen, and
Young In Yeo for helpful discussions about implementing and optimizing
CUDA code.  We thank Tom Loredo for bringing DEMCMC to our attention.
Finally, we thank nVidia Corp.\ for sponsoring AstroGPU
2007 and particularly Mark Harris and David Luebke for their contributions to
training astronomers in GPU programming.

\section*{References}

Belleman, R. G., Bedorf, J., Portegies Zwart, S. F.,  2008 High performance 
direct gravitational N-body simulations on graphics processing units II: An 
implementation in CUDA.  New Astronomy 13,  103-112.  

Danby, J.M.A 1988 Fundamentals of celestial mechanics.  Willman-Bell, Richmond.

Ford, E. B.,  2005 Quantifying the Uncertainty in the Orbits of Extrasolar 
Planets.  \aj 129,  1706-1717.  

Ford, E. B.,  2008 Adaptive Scheduling Algorithms for Planet Searches.  \aj 
135,  1008-1020.  

Ford, E. B., Gregory, P. C.,  2006 Bayesian Model Selection and Extrasolar 
Planet Detection.   arXiv:astro-ph/0608328- 

Gregory, P. C.,  2007 A Bayesian Kepler periodogram detects a second planet 
in HD208487.  \mnras 374,  1321-1333.  

Hamada, T., Iitaka, T.,  2007 The Chamomile Scheme: An Optimized Algorithm 
for N-body simulations on Programmable Graphics Processing Units.   
arXiv:astro-ph/0703100- 

Harris, C., Haines, K., Staveley-Smith, L.,  2008 GPU accelerated radio 
astronomy signal convolution.  Experimental Astronomy 22,  129-141.  

Kahan, W. 1965 Pracniques: further remarks on reducing truncation errors. Communications of the ACM 8, 40.

Loredo, T. J.,  2004 Bayesian Adaptive Exploration.  Bayesian Inference and 
Maximum Entropy Methods in Science and Engineering 707,  330-346.  

Moore, A. J., Quillen, A. C., Edgar, R. G.,  2008 Planet Migration through 
a Self-Gravitating Planetesimal Disk.   arXiv:0809.2855- 

Nyland, L., Harris, M., Prins, J. 2007 Fast N-Body Simulation with CUDA
in GPU Gems 3, ed. H. Nguyen, ch. 31, Addison-Wesley, 677-695.

nVidia Corporation. 2008. NVIDIA CUDA Compute Unified Device Architecture Programming Guide 2.0. \mbox{http://www.nvidia.com/cuda}

Portegies Zwart, S. F., Belleman, R. G., Geldof, P. M.,  2007 
High-performance direct gravitational N-body simulations on graphics 
processing units.  New Astronomy 12,  641-650.  

ter Braak, C.J.F. 2006 A Markov Chain Monte Carlo version of the
genetic algorithm Differential Evolution: easy Bayesian computing for
real parameter spaces.  Statistics and Computing 16, 239-249.

\newpage

%
\begin{table}
\label{Tab:Memory}
\caption{The memory usage for each GPU function.  Memory usage is measured in bytes.  Reads/writes are measured in floating point values. (4 bytes per single precision value, 8 bytes per double precision value)}
\begin{tabular}{lcccccc}
\hline
             & $M_{ijk}$ & $E_{ijk}$ & $\Delta v_{ijk}$ & $\chi^2_k$ & $\Delta v_{ijk}$ \& $\chi^2_k$ \\
\hline 
\multicolumn{6}{c}{single precision} \\
Registers/thread             & 11   & 20 & 27 & 15 & 32 \\
Local memory/thread         & 0    &  0 &  0 &  0 &  0 \\
Shared memory/block         & 1064 & 44 & 48 & 2100 & 2120 \\
Threads/block                       & 256 & 256 & 256 & 512 & 256 \\
\hline
\multicolumn{6}{c}{double precision} \\
Registers/thread                    & 27 & 44 & 64 & 32 & 64 \\
Local memory/thread                 & 0  & 152 & 304 & 304 &  304 \\
Shared memory/block                 & 2088 & 44 & 48 & 4160 & 4176 \\
Threads/block                       & 512 & 320 & 256 & 512  & 256 \\
\hline
\multicolumn{6}{c}{mixed precision} \\
Registers/thread                    & 28   & 20 & 27 & 15 & 32 \\
Local memory/thread                 & 0    &  0 &  0 &  0 &  0 \\
Shared memory/block                 & 2088 & 44 & 48 & 2100 & 2120 \\
Threads/block                       & 256 & 256 & 512 & 512 & 256 \\
\hline
\multicolumn{6}{c}{independent of precision} \\
Global reads/system   & \npl              & $(1+\nobs) \npl$ & $(4+\nobs) \npl$                                                                      & $\nobs \times \npl$ & $5\nobs \times \npl$ \\
Global reads/block    & \nobs & \nobs & 0 & 2\nobs & 2\nobs \\
Global writes/system  & $\npl \times \nobs$ & $\npl \times \nobs$   & $\npl \times \nobs$   & 1 & 1 \\
\hline
\end{tabular}
\end{table}

%
\begin{table}
\caption[]{Performance and accuracey for different implementations.
We set $\nobs=256$ and $\npl=4$, and draw both the orbital
period and the velocity amplitude from distributions uniform
in the log of the period and velocity over the intervals
$P\in[2{\mathrm d},10{\mathrm yr})$ and $K\in[1,500)\mps$.  The
remaining model parameters were drawn from uniform distributions:
$e\sim~U[0,99)$, $\omega\sim~U[0,2\pi)$, $M_0\sim~U[0,2\pi)$.  For
accuracy comparison (relative to CPU implementation) we use
$\nsys=30,720$ systems.  For GPU performance benchmarks, we set
$\nsys=122,880$. }
\label{Tab:Performance}
\begin{tabular}{lccc}
\hline
                               & single & double & mixed \\
\hline
\multicolumn{4}{c}{Kepler's Equation only} \\
Solutions per second (GPU)     & $4.19 \times 10^{9}$ & $1.80 \times 10^{8}$ & NA \\
Solutions per second (CPU)     & $3.54 \times 10^{6}$ & $3.26 \times 10^{6}$ & NA \\
\hline
\multicolumn{4}{c}{Full Model Evaluation ($\Delta v_{ijk}$ and $\chi^2_k$ in separate functions)} \\
Max. Fractional Error $\chi^2$ & 0.11          & $1.0 \times 10^{-4}$ & $1.3 \times 10^{-4}$ \\
Max. Absolute Error $\chi^2$   & $6.9 \times 10^{4}$  & 232           & 86 \\
L1 norm $\chi^2$               & $2.4 \times 10^{-4}$ & $4.4 \times 10^{-7}$ & $5.8 \times 10^{-7}$ \\
Systems per second (GPU)       & $2.1 \times 10^{6}$  & $1.1 \times 10^{5}$  & $1.0 \times 10^{6}$  \\ 
\multicolumn{4}{c}{Full Model Evaluation ($\Delta v_{ijk}$ and $\chi^2_k$ in single function)} \\
Max. Fractional Error $\chi^2$ & 0.11          & $1.0 \times 10^{-4}$ & $1.2 10^{-4}$ \\
Max. Absolute Error $\chi^2$   & $6.9 \times 10^{4}$  & 232           & 78 \\
L1 norm $\chi^2$               & $2.4 \times 10^{-4}$ & $4.4 \times 10^{-7}$ & $6.3 \times 10^{-7}$ \\
Systems per second (GPU)       & $1.4 \times 10^{6}$  & $1.1 \times 10^{5}$  & $8.6 \times 10^{5}$  \\
\multicolumn{4}{c}{Both implementations} \\
GPU-CPU Transfer Time/Compute Time & 6.9\%         & 1.4\%         & 7.7\% \\ 
Systems per second (CPU)       & $1.6 \times 10^{3}$  & $1.6 \times 10^{3}$  & NA  \\
\hline
\end{tabular}
\end{table}

\end{document}